\documentclass[conference]{IEEEtran}
\IEEEoverridecommandlockouts

\newcommand{\x}{\textbf{x}}
\newcommand{\mcM}{\mathcal{M}}

\renewcommand{\u}{\textbf{u}}

\usepackage{cite}
\usepackage{graphicx}  
\usepackage{amsmath} 
\usepackage{amssymb}
\usepackage{algorithm}
\usepackage{algorithmic}

\usepackage{array}
\begin{document}

\newtheorem{definition}{Definition}
\newtheorem{theorem}{Theorem}
\newtheorem{lemma}{Lemma}
\newtheorem{proposition}{Proposition}
\newtheorem{corollary}{Corollary}

\title{Adaptive Resource Allocation in Jamming Teams Using Game Theory*}

\author{

\IEEEauthorblockN{Ali Khanafer, Sourabh Bhattacharya and Tamer Ba\c{s}ar}

\IEEEauthorblockA{Coordinated Science Laboratory,\\ University of Illinois at Urbana-Champaign\\
\{khanafe2, sbhattac, basar1\}@illinois.edu}

{\thanks{*Research supported in part by grants from AFSOR and AFSOR MURI.}}

}

\maketitle

\begin{abstract}
In this work, we study the problem of power allocation and adaptive modulation in teams of decision makers. We consider the special case of two teams with each team consisting of two mobile agents. Agents belonging to the same team communicate over wireless ad hoc networks, and they try to split their available power between the tasks of communication and jamming the nodes of the other team. The agents have constraints on their total energy and instantaneous power usage. The cost function adopted is the difference between the rates of erroneously transmitted bits of each team. We model the adaptive modulation problem as a zero-sum matrix game which in turn gives rise to a a continuous kernel game to handle power control. Based on the communications model, we present sufficient conditions on the physical parameters of the agents for the existence of a pure strategy saddle-point equilibrium (PSSPE).
\end{abstract}

\section{Introduction}
The decentralized nature of wireless ad hoc networks makes them vulnerable to security threats. A prominent example of such threats is jamming: a malicious attack whose objective is to disrupt the communication of the victim network intentionally, causing interference or collision at the receiver side. Jamming attack is a well-studied and active area of research in wireless networks. Unauthorized intrusion of such kind has initiated a race between the engineers and the hackers; therefore, we have been witnessing a surge of new smart systems  aiming to secure modern instrumentation and software from unwanted exogenous attacks.

The problem under consideration in this paper is inspired by recent discoveries of jamming instances in biological species. In a series of playback experiments, researchers have found that resident pairs of Peruvian warbling antbirds sing coordinated duets when responding to rival pairs. But under other circumstances, cooperation breaks down, leading to more complex songs. Specifically, it has been reported that females respond to unpaired sexual rivals by jamming the signals of their own mates, who in turn adjust their signals to avoid the
interference \cite{tobias09}.


Ad hoc networks consist of mobile energy-constrained nodes. Mobility affects all layers in a network protocol stack including the physical layer as channels become time-varying \cite{GoldsmithWicker}. Moreover, nodes such as sensors deployed in a field or military vehicles patrolling in remote sites are often equipped with non-rechargeable batteries. Power control and adaptive resource allocation (RA) play, therefore, a crucial role in designing robust communications systems. At the physical layer, power control can be used to maximize rate or minimize the transmission error probability, see \cite{AzouziDebbah, BelmegaDebbah} and the references therein. In addition, in multi-user networks, power control can be used to regulate the interference level at the terminals of other users \cite{Scutari,WeiYu,ElBatt}. Due to the lack of a centralized infrastructure in ad hoc networks, distributed solutions are essential. In this work, similar to \cite{BelmegaDebbah,Scutari,WeiYu}, we model the power allocation problem as a noncooperative game, which allows us to devise a non-centralized solution. As a departure from previous research, however, the power control mechanism we propose splits the power budget of each player into two portions: a portion used to communicate with team-mates and a portion used to jam the players of the other team. More importantly, the objective function is chosen to be the difference between the cumulative bit error rate (BER) of each team; this allows for increased freedom in choosing physical design parameters, besides the power level, such as the size of modulation schemes.  

Adaptive RA mechanisms involve varying physical layer parameters according to channel, interference, and noise conditions in order to optimize a specific metric, such as spectral efficiency. Adapting the modulation scheme, choosing coding schemes, and controlling the transmitter power level are examples of adaptive RA schemes. Goldsmith and Chua demonstrated in \cite{GoldsmithChua} that adaptive RA provides five times more the spectral efficiency of nonadaptive schemes. In this work, we propose an adaptive modulation scheme based on a zero-sum matrix game played by both teams.   

The conflicting objectives of the two teams entails the use of game-theoretic framework to study this problem. We identify three main tasks that each team needs to perform: 
\begin{enumerate}
\item \emph{Optimal trajectory:} computing the optimal motion path for the agents
\item \emph{Power Allocation:} dividing power between internal communications and jamming
\item \emph{Adaptive Modulation:} choosing an appropriate modulations scheme
\end{enumerate}
We addressed Task 1 in \cite{gamecomm11} by posing the problem as a {\it{pursuit-evasion game}}. The optimal strategies of the players are obtained by using techniques from {\it{differential game theory}}. We also addressed Task 2 in \cite{gamecomm11} using \emph{continuous kernel static games}; we will generalize the formulation in this work to include a minimum rate constraint. This will lead to a more practical scheme as it guarantees a non-zero communications rate. Task 3 will be addressed in this work using \emph{static matrix games}. The saddle-point equilibrium of the power allocation problem is parametrized by the modulation schemes of the two teams. We therefore introduce a third game in order to arrive at the equilibrium modulation schemes. In fact, this gives rise to a \emph{games-within-games} structure: the optimal trajectory is first found, power allocation is then performed, and finally the optimal modulation is computed at each time instant. 

The main contributions of this paper are as follows. We introduce a third layer of games to our recent work \cite{gamecomm11} in order to perform adaptive modulation. We also generalize the power allocation problem introduced in \cite{gamecomm11} to ensure non-zero communications rate. We introduce an optimization framework taking into consideration constraints in energy and power among the agents. Moreover, we relate the problem of optimal power allocation for communication and jamming to the communication model between the agents. Finally, we provide a sufficient condition for existence of an optimal decision strategy among the agents based on the physical parameters of the problems.  

The rest of the paper is organized as follows. We formulate the problem in Section \ref{ProblemFormulation} and explain the underlying notation. The saddle-point equilibrium properties of the team power control problem are studied in Section \ref{PowerAllocation} with the specific example of systems employing uncoded M-quadrature amplitude modulations (QAM). We introduce our adaptive modulation scheme in Section \ref{AdaptiveModulation}. Simulation results are presented in Section \ref{Simulations}. We conclude the paper and provide future directions in Section \ref{Conclusion}.

\section{Problem Formulation} \label{ProblemFormulation}
Consider two teams of mobile agents. Each agent is communicating with members of the team it belongs to, and, at the same time, jamming the communication between members of the other team. In particular, each team attempts to minimize its own BER while maximizing the BER of the other team. We consider a scenario where each team has two members, though at a conceptual level our development applies to higher number of team members as well. Team A is comprised of the two players $\{1^{a},2^{a}\}$ and Team B is comprised of the two players $\{1^{b},2^{b}\}$. We assume that $f_{a}$ and $f_{b}$ are the frequencies at which Team A and Team B communicate, respectively, and $f_{a}\neq f_{b}$. 

Naturally, for an initial position $\x_{0}\in {\bf X}$, the outcome of the game $\pi$, is given by the difference in the BERs of both teams during the entire course of the game. Formally:

\begin{eqnarray*}
\pi(\x_{0},\u^{a}_{i},\u^{b}_{j})=N\cdot\int^{T}_{0}\underbrace{[p^{a}_{1}(t)+p^{a}_{2}(t)-p^{b}_{1}(t)-p^{b}_{2}(t)]}_{L}dt, 
\end{eqnarray*}
where $p^{a}_{i}(t)$ and $p^{b}_{j}(t)$ are the BERs of agent $i$ in Team A and and agent $j$ in Team B, respectively; $\u^a_i$ and $\u^b_j$ are likewise the control inputs of agents $i$ and $j$ in teams A and B, respectively; $N$ is the total number of transmitted bits which remains constant throughout the game; and $T$ is the time of termination of the game. We conclude that the objective of Team A is to minimize $\pi$, whereas that of Team B is to maximize it. 

Since the agents are mobile, there are limitations on the amount of energy available to each agent that is dictated by the capacity of the power source carried by each agent. The game is said to terminate when any agent runs out of power. Let $P^{a}_{i}(t)$ and $P^{b}_{j}(t)$ denote the instantaneous power for communication used by player $i$ in Team A and player $j$ in Team B, respectively. We model this restriction as the following integral constraint for each agent:
\begin{eqnarray}
\int^{T}_{0}P_{i}(t)dt\leq E
\label{eqn:encon}
\end{eqnarray}

For each transmitter and receiver pair, we assume the following communications model in the presence of a jammer which is motivated by \cite{poov}. Given that the transmitter and the receiver are separated by a distance $d$, and the transmitter transmits with constant power $P_{T}$, the received signal power $P_{R}$ is given by
\begin{eqnarray}
P_{R}=\rho P_{T}d^{-\alpha}, 
\label{eqn:power}
\end{eqnarray}
where $\alpha$ is the path-loss exponent and $\rho$ depends on the antennas' gains. Typical values of $\alpha$ are in the range of $2$ to $4$. According to the free space path loss model, $\rho$ is given by:
\begin{equation*}
\rho = \frac{G_TG_R\lambda^2}{(4\pi)^2},
\end{equation*}
where $\lambda$ is the signal's wavelength and $G_T$, $G_R$ are the transmit and receive antennas' gains, respectively, in the line of sight direction. In real scenarios, $\rho$ is very small in magnitude. For example, using nondirectional antennas and transmitting at $900$ MHz, we have $\rho = \frac{(1)(0.33)}{(4\pi)^2} = 6.896\times10^{-4}$.

The received signal-to-interference ratio (SINR) $s$ is given by
\begin{eqnarray}
s=\frac{P_{R}}{I+\sigma^2},
 \label{eqn:sinr}
\end{eqnarray}
where $\sigma^2$ is the power of the noise added at the receiver, and $I$ is the total received interference power due to jamming and is defined as in (\ref{eqn:power}).

Let $P^{a}_{i}(t)$ and $P^{b}_{j}(t)$ denote the instantaneous power levels for communication used by player $i$ in Team A and player $j$ in Team B, respectively. Since the agents are mobile, there are limitations on the amount of energy available to each agent that is dictated by the capacity of the power source carried by each agent. We model this restriction as the following integral constraint for each agent
\begin{eqnarray}
\int^{T}_{0}P_{i}(t)dt\leq E.
\label{eqn:encon}
\end{eqnarray}

In addition to the energy constraints, there are limitations on the maximum power level of the devices that are used onboard each agent for the purpose of communication. For each player, this constraint is modelled by $0\leq P^{a}_{i}(t),P^{b}_{i}(t)\leq P_{max}$. 

We also assume that players of each team have access to different M-QAM modulation schemes. We denote the set of available modulation sizes to the players in Team A by $\mcM^a$ and that available to players of Team B by $\mcM^b$. The sizes of the employed QAM modulation by the teams are M$^a \in \mcM^a$ and M$^b \in \mcM^b$. We assume that Team A can choose among $n$ different modulation schemes, and Team B chooses from a set of $m$ different schemes, i.e., $|\mcM^a| = n, |\mcM^b| = m$. 

The instantaneous BER depends on the SINR, the modulation scheme, and the error control coding scheme utilized. Communications literature contains closed-form expressions and tight bounds that can be used to calculate the BER when the noise and interference are assumed to be Gaussian \cite{Goldsmith}. For uncoded M-QAM, where Gray encoding is used to map the bits into the symbols of the constellation, the BER can be approximated by \cite{Palomar}
\begin{equation} \label{BERuncodedQAM}
p(t) = g(s) \approx  \frac{\zeta}{r} \mathcal{Q}\left(\sqrt{\beta s} \right),
\end{equation}
where $r=\log(M)$, $\zeta = 4(1 - 1/\sqrt{M})$, $\beta = 3/(M-1)$, and $\mathcal{Q}$(.) is the tail probability of the standard Gaussian distribution. 

To ensure a non-zero communication rate between the agents of each team, we impose a minimum rate constraint for each agent: $R^{a}_{i}(t),R^{b}_{i}(t)\geq \tilde{R}$, where $\tilde{R}>0$ is a threshold design rate, which we assume is the same for all agents. The results can be readily extended to networks of players having a different value of the minimum design rate.
 
\begin{figure}[t!]
\centering
\includegraphics[width=6cm]{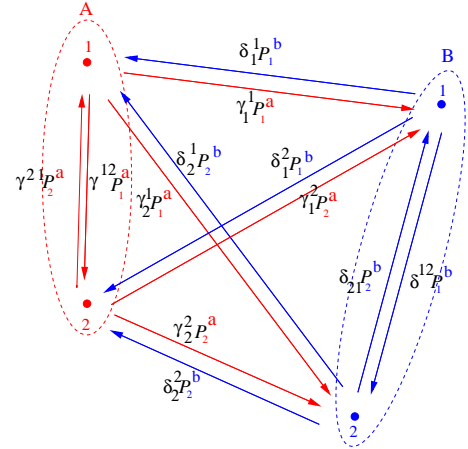}
\caption{Power allocation among the agents for communication as well as jamming. }
\label{fig:fig131}
\end{figure}

At every instant, each agent has to decide on the fraction of the power that needs to be allocated for communication and jamming. Table \ref{tbl:decvar} provides a list of decision variables for the players, which models the power allocation. Each decision variable is a non-negative real number and lies in the interval $[0,1]$. The decision variables belonging to each row add up to one. The fraction of the total power allocated by the player in row $i$ to the player in column $j$ is given by the first entry in the cell $(i,j)$. This allocated power is used for jamming if the player in column $j$ belongs to the other team; otherwise, it is used to communicate with the agent in the same team. Similarly, the distance between the agent in row $i$ and the agent in column $j$ is given by the second entry in cell $(i,j)$. Since distance is a symmetric quantity, $d^{ij}_{}=d^{ji}_{}$ and $d_{ij}^{}=d_{ji}^{}$. Fig. \ref{fig:fig131} depicts the power allocation between the members of the same team as well as between members of different teams.
\begin{table}[h]
\caption{Decision variables and distances among agents.}
\label {tbl:decvar}
\begin{center}
  \begin{tabular}{|c|c|c|c|c|}
\hline
&&&&\\
& $1^{b}$ & $2^{b}$&$1^{a}$&$2^{a}$\\
&&&&\\
\hline
&&&&\\
$1^{a}$&$\gamma^{1}_{1},d_{1}^{1}$&$\gamma^{1}_{2},d^{1}_{2}$&---&$\gamma^{12},d^{12}_{}$\\
&&&&\\
\hline
&&&&\\
$2^{a}$&$\gamma^{2}_{1},d_{1}^{2}$&$\gamma^{2}_{2},d_{2}^{2}$&$\gamma^{21},d_{}^{21}$&---\\
&&&&\\
\hline
&&&&\\
$1^{b}$&---&$\delta^{}_{12},d^{}_{12}$&$\delta^{1}_{1},d_{1}^{1}$&$\delta^{2}_{1},d_{1}^{2}$\\
&&&&\\
\hline
&&&&\\
$2^{b}$&$\delta^{}_{21},d^{}_{21}$&---&$\delta^{1}_{2},d_{2}^{1}$&$\delta^{2}_{2},d_{2}^{2}$\\
&&&&\\
\hline
\end{tabular}
\end{center}
\end{table}
In addition to power allocation, each team has to decide on the size of the QAM modulation to be used. In summary, each agent has to compute the following decision variables at each instance in accordance with the above tasks: (i) the instantaneous control (Task 1); (ii) the instantaneous power level, $P_{i}(t)$ (Task 2); (iii) all the decision variables present in the row corresponding to the agent in Table \ref{tbl:decvar} (Task 2); and (iv) the size of the QAM schemes, M$^a$ or M$^b$ (Task 3). 
\section{Power Allocation} \label{PowerAllocation}
From (\ref{eqn:sinr}), the received SINR and the rate achieved by each agent are given by the following expressions  
\begin{eqnarray}
s^{a}_{i} & = & \frac{\rho_aP^{a}_{j}(t)\gamma^{ji}_{}(d^{ij})^{-\alpha}}{\sigma^2+\rho_bP^{b}_{1}(t)\delta^{i}_{1}(d^{i}_{1})^{-\alpha}+\rho_bP^{b}_{2}(t)\delta^{i}_{2}(d^{i}_{2})^{-\alpha}} \nonumber \\
s^{b}_{i} & = & \frac{\rho_bP^{b}_{j}(t)\delta_{ji}(d_{ij})^{-\alpha}}{\sigma^2+\rho_aP^{a}_{1}(t)\gamma^{1}_{i}(d^{1}_{i})^{-\alpha}+\rho_aP^{a}_{2}(t)\gamma^{2}_{i}(d^{2}_{i})^{-\alpha}} \nonumber \\
R^a_i & = & \log(1+s^a_j), \quad R^b_i  =  \log(1+s^b_j)
\label{eqn:sinrexp}
\end{eqnarray}

Agents of the same team embark in a team problem which eliminates their need to exchange information about their decision variables. Further, since agents in different teams do not communicate, they possess information only about their own decision variables. This makes the power allocation problem a continuous kernel zero-sum game between the teams: 

\emph{Team A:} The objective of each agent is to minimize $L$.
\begin{equation} 
\min_{P^{a}_{i},\gamma^{i}_{1},\gamma^{i}_{2},\gamma^{ij}} L(M^a,M^b)\Rightarrow \min_{P^{a}_{i},\gamma^{i}_{1},\gamma^{i}_{2},\gamma^{12}} (\underbrace{p^{a}_{j}-p^{b}_{1}-p^{b}_{2}}_{L^{a}_{i}(M^a,M^b)})
\label{eqn:opt1a}
\end{equation}
subject to: \quad $0 \leq P^{a}_{i}(t)\leq P_{\text{max}}, \quad R^a_i \geq \tilde{R}$
\[\gamma^{i}_{1}+\gamma^{i}_{2}+\gamma^{ij}=1,\quad \gamma^{i}_{1},\gamma^{i}_{2},\gamma^{ij}\geq0\]

\emph{Team B:} The objective of each agent is to maximize $L$.
\begin{equation} 
\max_{P^{b}_{i},\delta^{1}_{i},\delta^{2}_{i},\delta_{ij}} L(M^a,M^b)\Rightarrow \max_{P^{b}_{i},\delta^{1}_{i},\delta^{2}_{i},\delta_{ij}} (\underbrace{p^{a}_{1}+p^{a}_{2}-p^{b}_{j}}_{L_{i}^{b}(M^a,M^b)})
\label{eqn:opt1b}
\end{equation}
subject to: \quad $0 \leq P^{i}_{b}(t)\leq P_{\text{max}},\quad R^b_i \geq \tilde{R}$
\[\delta^{1}_{i}+\delta^{2}_{i}+\delta_{ij}=1,\quad \delta^{1}_{i},\delta^{2}_{i},\delta_{ij}\geq0\]
Note that the power allocation vector for $1^a$ denoted by $\gamma = (\gamma^{12},\gamma^1_1,\gamma^1_2)$ belongs to the intersection between the three-dimensional simplex $\Delta^{3}$ and the plane $\mathbf{r}_0$, where $\mathbf{r}_0 = \{\gamma|\gamma^{12} = \frac{1}{a_1}(2^{\tilde{R}}-1)\}$. The power allocation vectors of other players belong to similar sets.

In \cite{gamecomm11}, we showed that the optimal value of the power consumption for each player is $P_{\max}$.  We also showed that the entire game terminates in a fixed time $T=\frac{E}{P_{\max}}$ irrespective of the initial position of the agents. Moreover we provided a sufficient condition for the existence of a pure-strategy saddle-point equilibrium (PSSPE) for the power allocation game when uncoded M-QAM schemes are used by all agents. Here, we modify the condition to allow teams to use different modulation schemes as made formal by the next theorem. 
\begin{theorem} \label{thm:QAMPSNE}
When all players employ uncoded $M$-QAM modulation schemes, the power allocation team game formulated above has a unique PSSPE solution if the following condition is satisfied:
\begin{equation}
P_{max} \cdot \max \left\{\frac{\rho_a(d^{12})^{-\alpha}}{M^a - 1}, \frac{\rho_b(d_{12})^{-\alpha}}{M^b - 1} \right\} < \sigma^2. 
\end{equation}
For the special case of $M^a = M^b = M$ and $\rho_b \approx \rho_a = \rho $, the condition becomes
\begin{equation}
\beta \rho P_{max} \left(\min\{d^{12}, d_{12} \}\right)^{-\alpha} < 3\sigma^2. \label{Thm4Cond}\\
\end{equation}
\end{theorem}
The proof is similar to that presented in \cite{gamecomm11} and is omitted here. Note that the left hand side of inequality (\ref{Thm4Cond}) depends entirely on physical design parameters; this is of particular importance for design purposes. Moreover, we showed in \cite{gamecomm11} that this condition can be expressed in terms of the received signal-to-noise-ratios (SNRs) for all players, which could be more insightful from a communication systems perspective. Consider, for example, 1$^a$, and let SNR$^x_y = \frac{P_{max} \gamma^x_y \rho  (d^x_y)^{-\alpha} }{\sigma^2}$ and SNR$_{xy} = \frac{P_{max} \delta_{xy} \rho  (d_{yx})^{-\alpha} }{\sigma^2}$. We then have:
\begin{eqnarray*}
\text{SNR}_{ij} & < & \frac{3}{\beta}(\text{SNR}_j^2 + 1)   \quad i,j\in\{1,2\}; i\neq j\\
\end{eqnarray*} 
Yet another useful way to interpret condition (\ref{Thm4Cond}) is regarding it as a minimum rate condition:
\begin{equation*}
r>\log\left(1+\frac{\rho P_{max} \left(\min\{d^{12}, d_{12} \}\right)^{-\alpha}}{\sigma^2}\right).
\end{equation*}

Assuming (\ref{Thm4Cond}) holds, the objective function is strictly convex in the decision variables of $1^a$, $2^a$ and strictly concave in the decision variables of $1^b$, $2^b$. A unique globally optimal solution $(\bar{\gamma})$ therefore exists, which we characterize using the KKT conditions \cite{Luenberger}. Consider, for example, the case of $1^a$. The expressions for SINR provided in (\ref{eqn:sinrexp}) relevant to the optimization problem being solved by $1^{a}$ can be written in a concise form as shown below:
\begin{eqnarray*}
s^{a}_{2}= a_{1}\gamma^{12},\quad
s^{b}_{1}=\frac{b_{1}}{c_{1}+\gamma_{1}^{1}},\quad
s^{b}_{2}=\frac{d_{1}}{e_{1}+\gamma_{2}^{1}},
\end{eqnarray*}
where 
\begin{eqnarray*}
a_1 & = & \frac{1}{\frac{\sigma}{P_{max}\rho_a(d^{12})^{-\alpha}}+\tilde{\rho}\delta^{2}_{1}\left(\frac{d^{2}_{1}}{d^{12}}\right)^{-\alpha}+\tilde{\rho}\delta^{2}_{2}\left(\frac{d^{2}_{2}}{d^{12}}\right)^{-\alpha}},\\
b_1 & = &\tilde{\rho}\delta_{21}\left(\frac{d_{12}}{d^{1}_{1}}\right)^{-\alpha}, \quad d_1 =  \tilde{\rho}\delta_{12}\left(\frac{d_{12}}{d^{1}_{2}}\right)^{-\alpha}, \\
c_1 & = & \frac{\sigma}{P_{max}\rho_a(d^{1}_{1})^{-\alpha}}+\gamma^{2}_{1}\left(\frac{d^{2}_{1}}{d^{1}_{1}}\right)^{-\alpha}, \\
e_1 & = & \frac{\sigma}{P_{max}\rho_a(d^{1}_{2})^{-\alpha}}+\gamma^{2}_{2}\left(\frac{d^{2}_{2}}{d^{1}_{2}}\right)^{-\alpha},\quad \tilde{\rho} = \left(\frac{\rho_b}{\rho_a}\right).
\end{eqnarray*}
The KKT conditions can then be written as:
\begin{eqnarray}
\nabla L_1^a(\bar{\gamma})+\displaystyle\sum^{4}_{i=1}\lambda_{i}\nabla h_{i}(\bar{\gamma})+\eta\nabla h(\bar{\gamma})=0,\\
\lambda_{i}h_{i}(\bar{\gamma})=0,\quad \lambda_{i},\eta\geq0,\quad i\in\{1,2,3\}\nonumber 
\end{eqnarray}
where 
\begin{eqnarray}
h_{1}(\bar{\gamma})&=&-\gamma^{12} + \min\left\{\frac{2^{\tilde{R}}-1}{a_1},1\right\} \leq 0 \nonumber\\
h_{2}(\bar{\gamma})&=&-\gamma^{1}_{1}\leq0, \quad h_{3}(\bar{\gamma})=-\gamma^{1}_{2}\leq0 \nonumber\\
h(\bar{\gamma})&=&\gamma^{12}+\gamma^{1}_{1}+\gamma^{1}_{2}-1=0\nonumber
\end{eqnarray}
Now, we present the necessary and sufficient conditions for the solution to the optimization problem for the agents. Let us consider the case of $1^{a}$. The assumptions in Theorem 4 regarding strict convexity of $L^{a}_{1}$ render the KKT conditions to be necessary as well as sufficient conditions for the unique global minimum. 

\begin{figure*}
$\mathbf{A}=$
\begin{align} \label{StaticMatrixGame}
\begin{tabular}{r|c|ccc|c|c}
\multicolumn{1}{r}{}
 &  \multicolumn{1}{c}{$\mcM^b(1)$}
 &  \multicolumn{1}{c}{$\mcM^b(2)$}
 & \multicolumn{1}{c}{...} 
 &  \multicolumn{1}{c}{$\mcM^b(m-1)$}
 &  \multicolumn{1}{c}{$\mcM^b(m)$} \\
\cline{2-6}
$\mcM^a(1)$ & $L(\mcM^a(1),\mcM^b(1))$ & $L(\mcM^a(1),\mcM^b(2))$ & ... & $L(\mcM^a(1),\mcM^b(m-1))$ & $L(\mcM^a(1),\mcM^b(m))$ \\
\cline{2-6}
$\mcM^a(2)$ & $L(\mcM^a(2),\mcM^b(1))$ & $L(\mcM^a(2),\mcM^b(2))$ & ...& $L(\mcM^a(2),\mcM^b(m-1))$ & $L(\mcM^a(2),\mcM^b(m))$\\
\vdots & \vdots & \vdots & \vdots & \vdots & \vdots & Team A \\
$\mcM^a(n)$ & $L(\mcM^a(n),\mcM^b(1))$ & $L(\mcM^a(n),\mcM^b(2))$ & ...& $L(\mcM^a(n),\mcM^b(m-1))$ & $L(\mcM^a(n),\mcM^b(m))$\\
\cline{2-6}
\multicolumn{1}{r}{}
 &  \multicolumn{1}{c}{}
 &  \multicolumn{1}{c}{}
 & \multicolumn{1}{c}{Team B} 
 &  \multicolumn{1}{c}{}
 &  \multicolumn{1}{c}{}
\end{tabular}
\end{align}
\hrulefill
\vspace{-0.3cm}
\end{figure*}

To this end, we obtain:
\[\nabla L_1^a=\left[\begin{array}{c}a_1g'(s^{a}_{2})\\ \frac{b_1g'(s^{b}_{1})}{(c_1+\gamma^{1}_{1})^{2}}\\ \frac{b_1g'(s^{b}_{2})}{(c_1+\gamma^{1}_{2})^{2}}\end{array}\right],\nabla h(\bar{\gamma})=\left[\begin{array}{c}1\\1\\1\end{array}\right]\]
\[\nabla h_{1}(\bar{\gamma})=\left[\begin{array}{c}-1\\0\\0\end{array}\right],\nabla h_{2}(\bar{\gamma})=\left[\begin{array}{c}0\\-1\\0\end{array}\right], \nabla h_{3}(\bar{\gamma})=\left[\begin{array}{c}0\\0\\-1\end{array}\right].
\]
Since $\gamma\in \Delta^{3} \cap \mathbf{r}_o$, at most three of the constraints can be active at any given point. Hence, the gradient of the constraints at any feasible point are always linearly independent.

If two of the three constraints among $\{h_{1},h_{2},h_{3}\}$ are active, then $\bar{\gamma}$ has a unique solution that is given by the vertex of the simplex that satisfies the two constraints. If only one of the constraints among $\{h_{1},h_{2},h_{3}\}$ is active, then we have the following cases depending on the active constraint
\begin{enumerate}
\item $h_{1}(\bar{\gamma}^{1})=0$: $\bar{\gamma}^{1}=(\tilde{\gamma}^{12},\gamma^{1*}_{1},1-\gamma^{1*}_{1}-\tilde{\gamma}^{12})$, where $\tilde{\gamma}^{12}  = \min\left\{\frac{2^{\tilde{R}}-1}{a_1},1\right\}$, satisfies the equation
\begin{equation}
g'(s_{2}^{b})\frac{d_{1}}{[e_{1}+(1-\gamma^{1*}_{1}-\tilde{\gamma}^{12})]^{2}}=g'(s_{1}^{b})\frac{b_{1}}{[c_{1}+\gamma^{1*}_{1}]^{2}}
\label{eqn:Case1}
\end{equation} 
\item $h_{2}(\bar{\gamma}^{2})=0$: $\bar{\gamma}^{2}=(1-\gamma^{1*}_{2},0,\gamma^{1*}_{2})$ satisfies the following equation
\begin{eqnarray}
a_{1}g'(s^{a}_{2})=\frac{d_{1}g'(s^{b}_{2})}{(e_{1}+\gamma^{1*}_{2})^{2}}
\label{eqn:Case2}
\end{eqnarray}
\item $h_{3}(\bar{\gamma}^{3})=0$: $\bar{\gamma}^{3}=(1-\gamma^{1*}_{1},\gamma^{1*}_{1},0)$ satisfies the following equation
\begin{eqnarray}
a_{1}g'(s^{a}_{2})=\frac{b_{1}g'(s^{b}_{1})}{(c_{1}+\gamma^{1*}_{1})^{2}}
\label{eqn:Case3}
\end{eqnarray}
\end{enumerate}
If none of the inequality constraints are active, then $\bar{\gamma}^{4}=(\underbrace{1-\gamma^{1*}_{1}-\gamma^{1*}_{2}-\tilde{\gamma}^{12}}_{\gamma^{12*}_{}},\gamma^{1*}_{1},\gamma^{1*}_{2})$ 
is the solution to:
\begin{eqnarray}
a_{1}g'(s^{a}_{2})-\frac{b_{1}}{[c_{1}+\gamma^{1*}_{1}]^{2}}g'(s^{b}_{1})=0\nonumber\\
a_{1}g'(s^{a}_{2})-\frac{d_{1}}{[e_{1}+\gamma^{1*}_{2}]^{2}}g'(s^{b}_{2})=0
\label{eqn:Case4}
\end{eqnarray}
Here, $\bar{\gamma}$ lies in the set $\{(1,0,0),(0,1,0),(0,0,1),\bar{\gamma}^{1},\bar{\gamma}^{2},\bar{\gamma}^{3},\bar{\gamma}^{4}\}$. An important point to note is that $a_{1},b_{1},c_{1},d_{1}$  and $e_{1}$ depend on the decisions of the other players. Therefore, the computation of the decision variables depend on the value of the decision variables of the rest of the players. A possible way to deal with this problem is to use iterative schemes for computation of strategies. \cite{Basar} provides some insights into the efficacy of such schemes from the point of view of convergence and stability. In this work, we assume that each agent has enough computational power so as to complete these iterations in a negligible amount of time compared to the total horizon of the game. The specific conditions for 1$^a$ corresponding to (\ref{eqn:Case1})-(\ref{eqn:Case3}) when M-QAM modulations are utilized are:
\begin{eqnarray}
\left(\frac{s_1^b}{s_2^b} \right)^{\frac{3}{2}}\exp\left(-\frac{\beta}{2}(s_1^b - s_2^b) \right) - \frac{b_1}{d_1} &=& 0 \\
\left(\frac{s_2^b}{s_2^a} \right)^{\frac{1}{2}}\exp\left(-\frac{\beta}{2}(s_2^a - s_2^b) \right) - \frac{a_1d_1}{(e_1+\gamma_2^1)^2} &=& 0 \label{eqn:2in1}\\
\left(\frac{s_1^b}{s_2^a} \right)^{\frac{1}{2}}\exp\left(-\frac{\beta}{2}(s_2^a - s_1^b) \right) - \frac{a_1b_1}{(c_1+\gamma_1^1)^2} &=& 0 \label{eqn:3in1}
\end{eqnarray}
Also, (\ref{eqn:Case4}) in this case corresponds to solving (\ref{eqn:2in1}) and (\ref{eqn:3in1}) jointly.
\section{Adaptive Modulation} \label{AdaptiveModulation}
The time-varying nature of the channels due to mobility emphasizes the need for robust communications. Adaptive modulation is a widely used technique as it allows for choosing the design parameters of a communications system to better match the physical characteristics of the channels in order to optimize a given metric such as: minimizing BER or maximizing spectral efficiency. In this work, we model the adaptive modulation as a matrix zero-sum game between the two teams. We therefore look for an equilibrium solution which would dictate what modulations should be adopted by the teams at each time instant. The competitive nature of the jamming teams makes our approach to the problem most practical as any other non-equilibrium solution cannot produce an improved outcome, relative to that yielded by the equilibrium, for any of the teams.

The matrix game is given in (\ref{StaticMatrixGame}). The rows are all the possible actions for players of Team A, and the columns are the different options available for Team B. The $(i,j)$-th element of the matrix is the value of the objective function $L$ when Team A employs $M^a = \mcM^a(i)$, and Team B employs $M^b = \mcM^b(j)$. 

A PSSPE does not always exist for the power allocation game. The condition for the existence of a PSSPE is $\min_i \max_j \mathbf{A}_{ij} = \max_j \min_i \mathbf{A}_{ij}$ \cite{Basar}. In case a PSSPE does not exist, we need to look for a solution in the larger class of mixed-strategies. A pair of strategies $\{M^{a*},M^{b*}\}$ is said to be a a mixed-strategy saddle point equilibrium (MSSPE) for the matrix game if \cite{Basar}
\begin{equation*}
(M^{a*})^T\mathbf{A}M^{b} \leq (M^{a*})^T\mathbf{A}M^{b*} \leq (M^{a})^T\mathbf{A}M^{b*} 
\end{equation*}
For an $n\times m$ matrix game, the following theorem from \cite{Basar}, which we state without proof, establishes the existence of an MSSPE for the adaptive modulation game.
\begin{theorem} \label{thm:AdaptiveMod}
The adaptive modulation game admits an MSSPE. 
\end{theorem}

In case multiple MSSPEs exist, the following corollary becomes essential \cite{Basar}.
\begin{corollary}
If $\{\mcM^a(i_1), \mcM^b(j_1)\}$ and $\{\mcM^a(i_2), \mcM^b(j_2)\}$ are two MSSPEs of the adaptive modulation game, then $\{\mcM^a(i_1), \mcM^b(j_2)\}$ and $\{\mcM^a(i_2), \mcM^b(j_1)\}$ are also MSSPEs.
\end{corollary}
This is termed the \emph{ordered interchangeability property} and its importance lies in that it removes any ambiguity associated with the existence of multiple equilibrium solutions as the teams do not need to communicate to each other which equilibrium solution they will be adopting. Literature contains different efficient low-complexity algorithms that computes MSSPEs for matrix games, such as Gambit \cite{Gambit}. We refer the interested reader to \cite{Basar} for a discussion of some of these approaches. Section \ref{Simulations} illustrates these concepts and shows how the choice of modulation size changes with the characteristics of the environment. Finally, it is assumed that players of each team communicate their modulation choices among themselves through a reliable side channel.

\section{Simulations Results} \label{Simulations}
To better understand the adaptive modulation scheme, we present the following example. Let $\mcM^a = \mcM^b = \{16,64,265 \}$, $d^1_1=17.7864$, $d^1_2=15.3376$, $d^2_1= 19.8951 $, $d^2_2=14.1128$, $d^{12}=20.6309$, $d_{12}=26.3224$, $\rho_a=0.0570$, $\rho_b= 0.0517$, $P_{max}=100$, $\sigma^2=10^{-3}$, $\tilde{R}=1$, and $\alpha = 2$. The matrix $\mathbf{A}$ corresponding to these values was found to be:
\begin{eqnarray*}
\mathbf{A} =  
\begin{tabular}{r|c|c|c|c}
\multicolumn{1}{r}{}
 &  \multicolumn{1}{c}{$16$}
 &  \multicolumn{1}{c}{$64$}
 &  \multicolumn{1}{c}{$256$} \\
\cline{2-4}
$16$ & $0.0158$ & $0.0533$ & $0.1229$ \\
\cline{2-4}
$64$ & $-0.0356$ & $0.0091$ & $0.0728$ & Team A\\
\cline{2-4}
$256$ & $-0.1155$ & $-0.0677$ & $0.0040$\\
\cline{2-4}
\multicolumn{1}{r}{}
 &  \multicolumn{1}{c}{}
 & \multicolumn{1}{c}{Team B} 
 &  \multicolumn{1}{c}{}
\end{tabular}
\end{eqnarray*}
Note that the third row dominates the other rows \emph{strictly}, and there is a unique PSSPE given by $\{256,256\}$ in this case. 

\begin{figure}[t!]
\centering
\includegraphics[width=8.5cm]{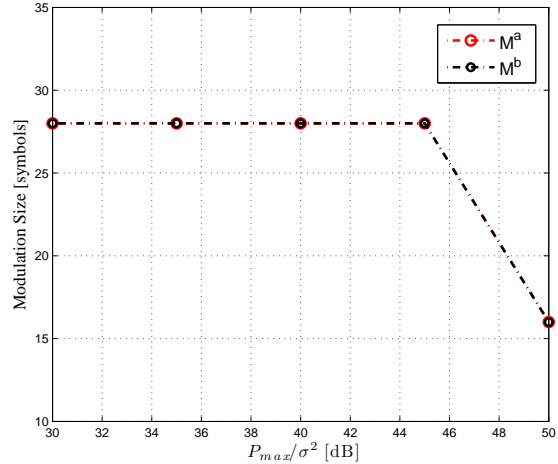}
\caption{Adaptive Modulation }
\label{fig:fig222}
\end{figure}

Fig. \ref{fig:fig222} depicts how the teams adapt their modulation scheme relative to the SNR, which we define as $P_{max}/\sigma^2$. The set of modulations available to each team is $\mcM = \{16,20,24,28\}$. Players $1^a$, $2^a$, and $1^b$ were placed close to each other, while $2^b$ is far from all of them; in particular: $d^1_1=2.2036$, $d^1_2=33.6830$, $d^2_1= 2.4211 $, $d^2_2=33.6393$, $d^{12}= 4.5607 $, $d_{12}= 33.2022 $. We also let $\rho_a=0.0570$, $\rho_b= 0.0517 $, $P_{max}=1$, $\tilde{R}=1$,  $\alpha = 3$, and varied the noise variance at all receivers to simulate the presented SNR range. We observe that both teams switch to a constellation of a smaller size at SNR$=50$ dB. This is due to both teams switching from pure communications to perform both communication and jamming. In order to do so, they both switch to a smaller constellation size which will guarantee robust communications for them as they will allocate some power to jam. 


\section{Conclusion and Future Work} \label{Conclusion}
This paper has studied the power allocation problem for jamming teams. An underlying static game was used to obtain the optimal power allocation, where the power budget of each user is split between communication and jamming powers. A separate matrix game was utilized in order to arrive at the optimal modulation schemes for each team. This work focused on the analysis of teams consisting of two players only; a potential future direction is to generalize the results to teams consisting of multiple agents. Moreover, future work will consider scenarios of players possessing incomplete information and study the problem in the context of Bayesian games.

\bibliographystyle{abbrv}
\bibliography{references}

\end{document}